\documentclass[12pt,preprint]{aastex}
\usepackage{ltgtsim}

\newcommand{\rb}{r_{\rm B}}
\newcommand{\tb}{t_{\rm B}}
\newcommand{\mdot}{\dot{M}}
\newcommand{\mdotb}{\dot{M}_{\rm B}}
\newcommand{\mdotbh}{\dot{M}_{\rm BH}}
\newcommand{\vinf}{v_{\infty}}
\newcommand{\rbh}{r_{\rm BH}}

\newcommand{\ra}{r_{\rm A}}
\newcommand{\calm}{\mathcal{M}}
\newcommand{\msun}{M_{\odot}}
\newcommand{\oms}{\omega_{*}}

\begin{document}

\title{Bondi Accretion in the Presence of Vorticity}

\author{Mark R. Krumholz}
\affil{Physics Department, University of California, Berkeley,
Berkeley, CA 94720}
\email{krumholz@astron.berkeley.edu}

\author{Christopher F. McKee}
\affil{Departments of Physics and Astronomy, University of California,
Berkeley, Berkeley, CA 94720}
\email{cmckee@astron.berkeley.edu}

\author{Richard I. Klein}
\affil{Astronomy Department, University of California, Berkeley,
Berkeley, CA 94720, and Lawrence Livermore National Laboratory,
P.O. Box 808, L-23, Livermore, CA 94550}
\email{klein@astron.berkeley.edu}

\begin{abstract}
The classical Bondi-Hoyle formula gives the accretion rate onto a
point particle of a gas with a uniform density and velocity. However,
the Bondi-Hoyle problem considers only gas with no net vorticity,
while in a real astrophysical situation accreting gas invariably has
at least a small amount of vorticity. We therefore consider the
related case of accretion of gas with constant vorticity, for the
cases of both small and large vorticity. We confirm the findings of
earlier two dimensional simulations that even a small amount of
vorticity can substantially change both the accretion rate and the
morphology of the gas flow lines. We show that in three dimensions the 
resulting flow field is non-axisymmetric and time dependent. The
reduction in accretion rate is due to an accumulation of circulation
near the accreting particle. Using a combination of simulations and
analytic treatment, we provide an approximate formula for the
accretion rate of gas onto a point particle as a function of
the vorticity of the surrounding gas.
\end{abstract}



\section{Introduction}

Accretion of a background gas by a point-like particle is a ubiquitous
phenomenon in astrophysics, occurring on size scales ranging from black
holes in galactic nuclei to individual stars or compact objects
accreting the winds of their companions. Bondi, Hoyle, and Lyttleton
\citep{hoyle39, hoyle40a, hoyle40b, hoyle40c, bondi52} treated this
situation by considering the ideal problem of a massive point particle 
within a uniform gas moving at a constant speed with respect to the
particle. For a polytropic gas, the problem is largely characterized
by a single dimensionless parameter, the Mach number $\calm$, which is
defined as the ratio of the gas-particle velocity to the sound speed
of the uniform gas. (The ratio of specific heats $\gamma$ also affects 
the accretion, but its effects are only order unity.) The Bondi-Hoyle
formula gives the approximate accretion rate as a function of $\calm$,
which, in a version updated by \citet{ruffert94a} and
\citet{ruffert94b}, is
\begin{equation}
\label{bhformula}
\mdotbh = 4 \pi \rho_{\infty} G^2 M^2 c_{\infty}^{-3} \left[
\frac{\lambda^2 + \calm^2}{\left(1+\calm^2\right)^4}
\right]^{1/2}
\end{equation}
Here, $\lambda$ is a number of order unity that depends on $\calm$ and 
on the equation of state. For $\calm=0$ (Bondi accretion) in an
isothermal medium, $\lambda=\exp(1.5)/4\approx 1.1$.

The Bondi-Hoyle formula finds wide application in astrophysics. Our
first motivation for exploring this topic is star formation. The
competitive accretion picture of star formation posits that
protostars form with initially low masses, typically $\ltsim
0.1\msun$, and grow in mass by accreting unbound gas from the
molecular clump in which they form \citep{bonnell1997, bonnell1998,
bonnell2001a, bonnell2001b, klessen2000, klessen2001}. The competition
for gas is invoked to explain the initial mass function (IMF). This is
a process of
Bondi-Hoyle accretion, albeit in a turbulent medium, and one ought to
be able to use a Bondi-Hoyle-like formula to estimate the rate
at which the proposed ``seed'' protostars would accrete. A second
significant area of application is in black holes, particularly
the supermassive black holes (SMBHs) at the centers of
galaxies. Numerous authors have used the Bondi formula to estimate
the rate at which SMBHs accrete, and found that it substantially
overestimates the accretion rate \citep{dimatteo01}. \citet{proga03}
have argued that rotation may explain the difference between accretion
rates estimated from observations and those predicted by the Bondi
rate. A third area of application is in accretion rates onto compact
objects. \citet{perna03} suggest that a combination of rotation and
magnetic fields leads to lower accretion rates onto isolated neutron
stars than one would estimate using the Bondi formula. They argue that
this lowered accretion rate accounts for the unexpectedly small number
of isolated neutron stars that have been detected. 

Previous authors have considered the problem of hydrodynamic accretion with
non-zero vorticity, both analytically \citep{sparke80, sparke82,
abramowicz81} and numerically \citep{fryxell88, ruffert97, ruffert99,
igumenshchev00a,
igumenshchev00b, proga03}. \citet{abramowicz81} analytically treated
accretion of gas with non-zero angular momentum onto a black
hole. They found that for very low angular momentum the flow pattern
is similar to Bondi
accretion; the flow is quasi-spherical, it becomes sonic at a radius
much larger than the Schwarzschild radius, and the accretion rate is
comparable to the Bondi rate. For high angular momentum, the gas forms 
a disk and the sonic radius in the equatorial plane becomes only a
factor of a few times the Schwarzschild radius. The accretion rate is
substantially below the Bondi rate. \citet{proga03} performed two
dimensional axisymmetric simulations of the accretion of slowly
rotating gas onto a black hole. They find three regimes of
accretion. For sufficiently low total angular momentum, the flow is
Bondi-like,
similar to the pattern \citet{abramowicz81} predicted. For
intermediate total angular momentum, the highest angular momentum gas
that cannot accrete forms a dense torus about the black hole. The
torus blocks off accretion except through a thin funnel along the
poles, reducing the accretion rate to a value roughly independent of
the total angular momentum. Finally, for even higher total angular
momentum, a centrifugal barrier, coupled with the dense torus,
further reduces the accretion rate.

Our goal in this paper is to unify this work by describing the
accretion process using a one-parameter family similar to the
Bondi-Hoyle-Lyttleton treatment, and to give an analogous formula for
the accretion rate as a function of this parameter. We do this using a
combination of three-dimensional simulations and analytic
treatment. In these two important respects, our work is more complete
than previous work. In \S\ref{analytic}
we give an analytic analysis of the problem, and present our
approximate formula. In \S\ref{method} we describe the computational
methodology we use in our simulations, and in \S\ref{simulations} we
compare the results of our simulations to the analytic theory we put
forth. We summarize our conclusions in \S\ref{conclusions}.

\section{Analytic Treatment}

\label{analytic}

We consider a point particle accreting in a medium with non-zero
vorticity. The simplest case of this, which we treat here, is a medium 
with constant density and vorticity far from the particle, and where
the gas has no net velocity of relative to the particle. Thus, we
consider a particle at the origin of our coordinate system, surrounded
by a gas whose velocity far from the particle is
\begin{equation}
\mathbf{\vinf} = \oms c_{\infty} \frac{y}{\rb} \hat{x},
\end{equation}
where $\rb\equiv GM/c_{\infty}^2$ is the standard Bondi radius, $M$ is 
the mass of the particle,
$c_{\infty}$ is the sound speed of the gas at infinity, and $y$ is the 
$y$ coordinate. The factor
$\oms$ is a dimensionless number which we call the vorticity
parameter, since the vorticity of the velocity field is $-\oms
c_{\infty}/\rb \hat{z}$, independent of position.

The significance of $\oms$ is clear: $\oms=1$ corresponds to the
case where gas with an impact parameter $y=\rb$ is traveling at the sound
speed at infinity. The Keplerian velocity at a distance of $\rb$
from the central object is
\begin{equation}
v_K(\rb) = \sqrt{\frac{GM}{\rb}} = c_{\infty}.
\end{equation}
Thus, for $\oms=1$, gas whose impact parameter is $\rb$
is traveling at the Keplerian velocity for that impact
parameter. Since $v_K\propto r^{-1/2}$, while for our initial velocity
field $\vinf \propto y$, gas with smaller impact parameters is
sub-Keplerian, while gas with larger impact parameters is
super-Keplerian. For $\oms < 1$, the transition from
sub- to super-Keplerian $\vinf$ occurs at impact parameters
larger than $\rb$; for $\oms > 1$, the transition occurs at impact
parameters smaller than $\rb$. Because the Bondi radius is the distance at 
which gas comes within the gravitational sphere of influence of the
central particle, provided its kinetic energy at infinity is small, we
should expect a transition in behavior at $\oms \approx 1$. 

We note that previous authors (e.g. \citet{abramowicz81, proga03})
have usually parameterized their flows in terms of specific angular
momentum rather than vorticity, as we do here. The choice of
parameterization is somewhat a matter of taste. Specific angular
momentum has the advantage that it yields a circularization radius for
the flow that depends only on the specific angular momentum,
simplifying the analysis. On the other hand, the choice of vorticity
enables us to invoke the Kelvin Circulation Theorem in our
analysis, which we find helpful. In addition, since specific angular
momentum is a function of accretor position, and gas farther than
$\rb$ from an accretor is unaware of its presence, arranging a flow
with constant specific angular momentum requires an unlikely
coincidence. It is more natural to characterize a flow by its velocity
gradient, hence its vorticity, which is independent of accretor position.
Regardless of
the choice of parameter, however, it is possible to make an
approximate translation of the formulae we derive based on $\oms$ to
ones in terms of the specific angular momentum. For the initial
conditions we are considering, the specific angular momentum of a gas
parcel with impact parameter $y$ is
\begin{equation}
\label{linf}
l_{\infty} = y \vinf = \oms c_s \frac{y^2}{\rb}.
\end{equation}
In the artificial case of a flow with constant specific angular
momentum rather than constant vorticity, the equivalent value of
$\oms$ should be determined roughly by the value of the vorticity at
the Bondi radius. Thus, as an approximate conversion we suggest $\oms
= l_{\infty}/(c_s \rb)$.

Before moving on to analyze the behavior of accretion as a function of 
$\oms$, we consider briefly the question of the gas equation of state. We
choose to restrict ourselves to considering an isothermal
gas, and we denote the constant sound speed as $c_s$. This is physically
appropriate for the star formation context: radiative energy transport
in the dense molecular gas from which stars form is capable of
eliminating any temperature variations on times small compared to the
mechanical time scales in the gas \citep{semadeni00}. In the context of
SMBHs and compact objects, the thermodynamics is more complicated, as
the gas can range from isothermal to adiabatic depending on the
efficiency of radiative energy transport. However, based on prior
simulation work with Bondi accretion \citep{ruffert94a,ruffert94b}, and 
accretion with shear \citep{ruffert97}, we do not believe that changes
in the equation of state will change either the flow pattern or the
accretion rate by more than a factor of order unity.

\subsection{Very Small $\oms$}

\label{verysmall}

\citet{abramowicz81} performed the first detailed analysis of the
cases of small and very small $\oms$ (which we define below),
that we shall review and extend. We can distinguish two regimes that
occur for $\oms \ll 1$, depending on the physical dimensions of the
accreting object. In the absence of viscosity or other mechanisms of
angular momentum transport, a parcel of gas has constant specific
angular momentum $l_{\infty}$ as it falls towards the
accretor. The orbit of the gas parcel changes from predominantly
radial to predominantly tangential at a characteristic circularization 
radius, where the parcel's velocity is roughly the Kepler
velocity. The specific angular momentum of a parcel of gas in
Keplerian rotation at a distance $r$ from the central object is
\begin{equation}
\label{lkep}
l_K(r) = r v_K(r) = \sqrt{GMr},
\end{equation}
and the circularization radius is therefore determined by the condition
$l_{\infty} = l_K(r_{\rm circ})$. Equating (\ref{linf}) and (\ref{lkep}),
a gas parcel at initial impact parameter $y$ has a circularization radius
\begin{equation}
r_{\rm circ} = \oms^2 \frac{y^4}{\rb^3}.
\end{equation}
If the circularization radius for a given parcel of gas is small
compared to the radius of the accretor, $\ra$, then that gas will
reach the surface of the accretor traveling in a predominantly radial
orbit. Its angular momentum, and its vorticity, will therefore be
deposited in the central object and will not otherwise affect the flow
pattern, leading to an accretion rate approximately equal to the Bondi
rate.

In the case of small vorticity, where the velocity of the gas at
infinity is small compared to the Keplerian velocity, gas begins to
feel the influence of the accretor at a characteristic radius of
$\rb$. We can therefore distinguish our two regimes based on the
relative sizes of $r_{\rm circ}(\rb)$ and $\ra$. If $r_{\rm circ}(\rb)
< \ra$, then we are in the case analogous to that of a small
circularization radius in a flow with constant specific angular
momentum: the gas that reaches the
accretor will be on primarily radial orbits, and the accretion rate
should be the Bondi rate. The condition for this to occur is
\begin{equation}
\oms < \omega_{\rm crit} \equiv \sqrt{\frac{\ra}{\rb}}.
\end{equation}
We term the case $\oms < \omega_{\rm crit}$ ``very small $\oms$.''
To get a sense of values of $\omega_{\rm crit}$ in typical cases,
consider some of the applications we discussed in the introduction. In 
a star forming region, a typical accretor mass is $M\sim \msun$ and a
typical sound speed is $c_s\sim 0.2$ km s$^{-1}$, giving a Bondi radius
of $\rb \sim 0.1$ pc. A typical protostellar radius is $\ra \sim
10^{12}$ cm, so $\omega_{\rm crit}\sim 10^{-3}$. For the SMBH at the
center of the Milky Way, $M \approx 3.7 \times 10^6\msun$ for the
standard distance estimate \citep{ghez04}. The temperature is
not constant and depends on the details of radiative processes, but a
reasonable estimate for gas far from the black hole is $\sim 10^7$ K
\citep{melia94}, giving $c_{\infty}\sim \mbox{few} \times 10^7$ cm
s$^{-1}$
and $\rb\sim 0.1$ pc. The accretor radius is the Schwarzschild radius 
$\ra=2GM/c^2\sim 10^{12}$ cm, so again $\omega_{\rm crit}\sim
10^{-3}$.

Note that this condition $\oms\ltsim \omega_{\rm crit}$ for Bondi-like
accretion is roughly equivalent to the conditions of
\citet{abramowicz81} and \citet{proga03} on the specific angular
momentum when the accreting object is a non-rotating black hole. In
that case, the accretor radius is the Schwarzschild radius, $\ra =
2GM/c^2$, so $\omega_{\rm crit} = \sqrt{2c_s/c}$, and the specific
angular momentum of gas with impact parameter $y=\rb$ is
\begin{equation}
l_{\infty}(\rb) = \omega_{\rm crit} c_s \rb = \sqrt{2} \ra c.
\end{equation}
In their analytical treatment, \citet{abramowicz81} estimate that the
critical specific angular momentum for Bondi-like accretion
$l_{\infty}=2\ra c$; however, the exact critical value will depend on
the details of how the accretion flow joins onto the central object.

\subsection{Small $\oms$}

If the vorticity is larger than $\omega_{\rm crit}$, but still small
compared to unity, then gas circularizes before it reaches the accretor. The
accretor is effectively a point particle, and cannot absorb angular
momentum or vorticity from the flow. In this case, \citet{proga03}
have found in their simulations that gas with too much angular
momentum to accrete builds up into a torus around the accretor. This
thick torus inhibits accretion by blocking off a large part of the
solid angle through which streamlines could otherwise reach the
accretor.

We can explain the growth of the torus by making a useful analogy
to magnetohydrodynamics. The flow at infinity contains a constant
vorticity. In regions of the flow that are non-viscous, the
Kelvin Circulation Theorem requires that the vorticity contained in
an area element of the fluid is constant. This is analogous to
flux-freezing in a magnetic medium. The accretor's gravity drags lines 
of vorticity in toward the accreting particle. Near the particle,
where viscosity may become significant (for example due to
magneto-rotational instability in the accretion disk), gas can slip
through the vortex lines and reach the accreting object. However, as
long as the region of significant viscosity is confined near the
accretor, the vortex lines cannot escape. As accretion drags gas
inward, vortex lines build up near the accretor, just as magnetic flux
would build up in a magnetic flow.

The effect of this vorticity build-up is to increase the rotational
velocity in the gas. Consider a circular curve $C$ of radius $r_C$ in
the $xy$ plane, centered at the origin. The circulation around curve
$C$ is the line integral of the velocity,
\begin{equation}
\Gamma = \oint_C \mathbf{v}\cdot \mathbf{dl} = r_C \int v_{\phi}
d\phi,
\end{equation}
and by Stokes' theorem
\begin{equation}
\Gamma = \int_A \mbox{\boldmath$\omega$\unboldmath} \cdot \hat{z}\; dA,
\end{equation}
where \boldmath $\omega$ \unboldmath is the vector vorticity at any
point in the flow and $A$ is the area inside curve $C$. As a reference
state, consider a velocity field with $\mathbf{v}=\mathbf{v}_{\infty}$
everywhere. The
accretion process increases vorticity within the curve relative to this
reference state, so it must also increase $\Gamma$ and hence the
mean rotational velocity $v_{\phi}$
around $C$. Because the vorticity parameter $\oms<1$, we know that
initially $v_{\phi} < v_K$ as long as $r_C<\rb$. However, this
analysis shows that $v_{\phi}$ increases with time, so eventually the
rotational velocity will become comparable to the Keplerian velocity
and a rotationally supported disk must form. The Keplerian
velocity is supersonic for all radii smaller than $\rb$, and the
inflow onto the disk is not purely radial, so the disk will likely
develop internal shocks and a time-dependent structure.

The process of vorticity increase near the accretor cannot continue
indefinitely. Eventually, inward transport of vorticity by accreting
matter must be balanced by an outward flow of vorticity in gas leaving 
the vicinity of the accretor and escaping to infinity. The radius at
which gas becomes unbound from the accretor and can start escaping is
$\rb$, and we therefore expect that the disk will extend out to
$\sim\rb$ in radius. The scale height of a thin Keplerian disk is
\begin{equation}
H = \frac{c_s}{\Omega}
= c_s \sqrt{\frac{r^3}{GM}}
= r \sqrt{\frac{r}{\rb}},
\end{equation}
where $\Omega$ is the angular velocity. Since the disk
radius is $\sim\rb$, we expect $H\sim \rb$. Thus, the scale height is
comparable to the radius and we have, rather than a disk, a thick
torus. The torus may be somewhat thicker for an equation of state with
ratio of specific heats $\gamma$ greater than unity, because in that case the
sound speed will increase inside the Bondi radius due to compressional 
heating. However, the compression should be is only order unity at
$r\sim\rb$, so the effect of compression on the torus scale height
should also be order unity.

The circulation trapped within the torus should be
\begin{equation}
\Gamma \sim 2\pi\rb v_K(\rb) = 2\pi c_s\rb.
\end{equation}
In our reference state with $\mathbf{v}=\mathbf{v}_{\infty}$
everywhere, the circulation is
\begin{equation}
\Gamma_0 = \pi \rb^2 \omega = \pi c_s \rb \oms,
\end{equation}
which is clearly much smaller than the equilibrium value of $\Gamma$
for $\oms\ll 1$. The time required for the circulation to build up to
$\Gamma$ is determined by the rate at which mass falling onto the
accreting object carries vorticity inwards. The gas within the Bondi
radius reaches the central object, or the growing disk, on the Bondi
time scale, $\tb \equiv \rb/c_s$. Thus, the flow replaces the gas within
the Bondi radius on this time scale. Each set of ``replacement'' gas
carries the same vorticity as the gas it is replacing, and therefore
increases the circulation by $\Gamma_0$. As a result, the time
required for the accumulating vorticity to reach its equilibrium value
is approximately
\begin{equation}
t_{\rm eq} \sim \frac{\Gamma}{\Gamma_0} \tb = \frac{2}{\oms} \tb.
\end{equation}
This estimate only applies for $\oms\ltsim 1$, when
accumulation of circulation is the key factor in setting the final
accretion rate.

We can roughly estimate the amount by which the accretion rate
will be reduced by the presence of the torus once the flow reaches
equilibrium. The torus reduces accretion by blocking off solid angle
through which material could 
otherwise accrete. If the torus extends out to $\rb$, and its height
is roughly $\rb$ as well, it blocks all angles in the range $\pi/4 <
\theta < 3\pi/4$. If the reduction in accretion rate is proportional
to the fraction of solid angle blocked, then the accretion rate for
this case should be reduced to $\sim 30\%$ of the Bondi rate,
$\mdotb$, given by equation (\ref{bhformula}) for $\calm=0$.

It is important to note that the equilibrium state we describe here
does not depend on the value of $\oms$, as long as $\omega_{\rm
crit} < \oms \ll 1$. The equilibrium radius of the torus depends only
on the maximum vorticity that can accumulate near the accretor before
the velocity becomes comparable to the Kepler velocity and halts
infall in the disk. Thus, the equilibrium radius of the torus is $\sim
\rb$ independent of $\oms$, and the scale height and accretion rate
are also $H\sim\rb$ and $\mdot\sim 0.3\mdotb$,
respectively. Different values of $\oms$ only mean that vorticity
builds up more or less slowly, and thus that there is a longer or
shorter initial transient.

\subsection{Large $\oms$}

The final case to consider is $\oms\gtsim 1$. In this regime, the
velocity of the gas at the Bondi radius is comparable to or larger
than the Keplerian velocity. To derive an approximate accretion rate,
we adopt an ansatz from Bondi-Hoyle accretion.

Regardless of its velocity, gas
at infinity initially has a positive total energy. Its potential
energy is zero and it has non-zero internal energy. In order to
become bound and accrete onto the point mass, the gas must lose its
energy by shocking. If gas does shock, it is likely to accrete. Since
the flow at infinity is laminar, shocks will only occur for those
streamlines that are bent significantly by the gravity of the
accreting particle. To determine if a straight streamline is
susceptible to bending, we can compare the velocity of gas on that
streamline (adding both bulk and thermal velocity) to the maximum value of
the escape velocity along that streamline. If the peak escape
velocity is larger, then the streamline will bend and is likely to
terminate in a shock. Suppose the gas at infinity is traveling in the 
$x$ direction, and consider a streamline that begins at $xyz$
coordinates $(\pm\infty, y, z)$. The streamline will bend
significantly and pass through a shock if
\begin{equation}
\label{vesccondition}
\frac{2GM}{\sqrt{y^2+z^2}} > \vinf(y,z)^2+c_s^2.
\end{equation}
In the Bondi-Hoyle case, $\vinf=\calm c_s$ independent of $y$ and
$z$. Thus, (\ref{vesccondition}) reduces to
\begin{equation}
\sqrt{y^2+z^2} < \frac{2GM}{c_s^2 (1+\calm^2)} \equiv 2 \rbh,
\end{equation}
where we have used the standard definition for the Bondi-Hoyle
radius.

If we make the assumption that all parcels of gas traveling on
streamlines that bend significantly go through shocks, and all gas
that shocks loses enough energy to become bound and eventually reach
the accretor, then the rate at which mass accretes onto the particle
is equal to the rate at which gas from infinity begins traveling on
streamlines that meet condition (\ref{vesccondition}). We can therefore
define an area $A$ in the plane at infinity that satisfies
(\ref{vesccondition}) and approximate the accretion rate by
\begin{equation}
\mdot \approx \int_{A} \rho_{\infty} \sqrt{\vinf(y,z)^2+c_s^2}\; dy\, dz,
\end{equation}
where $\rho_{\infty}$ is the density of the gas at infinity and we
have again added the bulk and thermal velocities of the gas in
quadrature. For the Bondi-Hoyle case, $A$ is simply a circle of radius 
of radius $2\rbh$ and $\vinf$ is a constant, so the integral is
trivial. Evaluating it gives
\begin{equation}
\mdotbh \approx 4 \pi \rbh^2 \rho_{\infty} c_s \sqrt{1+\calm^2}
= 4 \pi \rho_{\infty}
\frac{(G M)^2}{c_s^3 \left(1+\calm^2\right)^{3/2}},
\end{equation}
which is just the Bondi-Hoyle formula (equation \ref{bhformula}) up
to factors of order unity.

Because this ansatz allows us to reproduce the Bondi-Hoyle formula, we
will apply it to predict the case of accretion with vorticity. In
this case, $\vinf=\oms c_s y/\rb$, so (\ref{vesccondition}) becomes
\begin{equation}
\frac{2GM}{\sqrt{y^2+z^2}} >
	\left(\oms \frac{c_s}{\rb}\right)^2 y^2+c_s^2.
\end{equation}
With the convenient change of variable $y'=y/\rb$ and $z'=z/\rb$, this 
becomes
\begin{equation}
2 > \left(y'^2+z'^2\right)^{1/2} \left(1+\oms^2 y'^2\right).
\label{Aomega1}
\end{equation}
We can solve (\ref{Aomega1}) for $z'$ to find the function
$Z'(y',\oms)$ that defines the boundary:
\begin{equation}
Z'(y',\oms) = \frac{\sqrt{4-y'^2(1+\oms^2 y'^2)^2}}{1+\oms^2 y'^2}.
\label{Aomega}
\end{equation}
We refer to region of the $yz$ plane bounded by $\pm Z'(y',\oms)$
as $A(\oms)$. The maximum value of $y'$ on the boundary of $A(\oms)$
occurs for $z'=0$, and is
\begin{equation}
y'_{\rm max}(\oms) =
\frac{3^{1/3} \left(9\oms + \sqrt{3+81\oms^2}\right)^{2/3} - 3^{2/3}}
{3 \oms \left(9\oms+\sqrt{3+81\oms^2}\right)^{1/3}}.
\end{equation}
Figure \ref{Aplot} shows the boundary of $A(\oms)$ as a
function of $\oms$.

The approximate accretion rate is
\begin{eqnarray}
\mdot(\oms) & \approx &
	\rb^2 \rho_{\infty} c_s
	\int_{A(\oms)} \sqrt{1+\oms^2 y'^2} \; dy' \; dz' \\
& \equiv & 4\pi \rb^2 \rho_{\infty} c_s f(\oms),
\label{mdotapprox1}
\end{eqnarray}
where we have defined the numerical factor
\begin{eqnarray}
f(\oms)
& = & \frac{1}{4\pi} \int_{A(\oms)} \sqrt{1+\oms^2 y'^2} \; dy' \; dz' \; \\
& = & \frac{1}{\pi}\int_0^{y'_{\rm max}(\oms)} \sqrt{1+\oms^2 y'^2}
	\int_0^{Z'(y',\oms)} dz' \; dy' \\
& = & \frac{1}{\pi}\int_0^{y'_{\rm max}(\oms)}
	\sqrt{\frac{4-y'^2 (1+\oms^2 y'^2)^2}{1+\oms^2 y'^2}} dy'.
\label{fomega}
\end{eqnarray}
The integral is straightforward to evaluate numerically, and we plot
the result in Figure \ref{fplot}. We can also obtain an approximation
for $f(\oms)$ in the limit $\oms\rightarrow\infty$. We show in
Appendix \ref{integralapprox} that in the limit $\oms\gg 1$, we can
approximate $f(\oms)$ by
\begin{equation}
\label{fapprox}
f(\oms) \approx \frac{2}{3\pi\oms} \ln\left(16\oms\right).
\end{equation}
We also show this approximation in Figure \ref{fplot}.

We must make one further modification to our estimated accretion
rate. In the limit $\oms\rightarrow 0$, $f(\oms)\rightarrow 1$
and (\ref{mdotapprox1}) reduces to approximately
$\mdotb$. Thus, this approximation smoothly interpolates between the
cases $\oms\ll 1$ and $\oms\gtsim 1$. However, we have already
seen that even in the case $\oms\ll 1$ the accretion rate can be
substantially below the Bondi rate, since
the accumulation of circulation leads to the formation of a thick torus
that blocks streamlines from reaching the accretor. The same
phenomenon should happen with $\oms\gtsim1$. Circulation will
still accumulate near the accretor. The shape of the torus should be the
same as in the $\oms\ll 1$ case, because that is set just by the physics
of disks. We therefore reduce our estimated accretion rate
(\ref{mdotapprox1}) by a constant factor so that it matches our
estimate of $0.3\mdotb$ for the case $\omega_{\rm crit}<\oms\ll
1$. The required factor is $0.34$; it is not exactly 0.3 because
(\ref{mdotapprox1}) for $\oms\ll 1$ differs from the Bondi
rate by a small factor.

With this modification in place, (\ref{mdotapprox1}) should give a
good approximation of the accretion rate for all $\oms>\omega_{\rm
crit}$. Our final estimate for the accretion rate by a flow with
vorticity $\oms$ at infinity is
\begin{equation}
\label{mdotapprox}
\mdot(\oms) \approx
4\pi \rho_{\infty} \frac{(GM)^2}{c_s^3}\cdot
\left\{ \begin{array}{r@{\quad:\quad}l}
\exp(1.5)/4 & \oms<\omega_{\rm crit} \\
0.34\, f(\oms) & \oms > \omega_{\rm crit}
\end{array}.
\right.
\end{equation}
In \S\ref{conclusions} we give a more accurate approximation that is
calibrated by our simulations.

\section{Computational Methodology}

\label{method}

To test the theory presented in the previous section, we ran a series
of simulations. In this section, we describe the simulation
methodology, and in the next we compare the simulation results to our
theory.

\subsection{Code}

The calculations in this paper use our three dimensional adaptive
mesh refinement (AMR) code to solve the Euler equations of
compressible gas dynamics
\begin{eqnarray}
\frac{\partial\rho}{\partial t} + \nabla\cdot\left(\rho
\mathbf{v}\right) & = & 0 \\
\frac{\partial}{\partial t} \left(\rho \mathbf{v}\right) + \nabla
\cdot \left(\rho \mathbf{vv}\right) & = & -\nabla P -
\rho\nabla\phi \\
\frac{\partial}{\partial t}\left(\rho e\right) + \nabla \cdot
\left[\left(\rho e + P\right)\mathbf{v}\right] & = & \rho
\mathbf{v} \cdot \nabla\phi,
\end{eqnarray}
where $\rho$ is the density, $\mathbf{v}$ is the vector
velocity, $P$ is the thermal pressure (equal to $\rho c_s^2$ since we
adopt an isothermal equation of state), $e$ is the total
non-gravitational energy per unit mass, and $\phi$ is the gravitational 
potential. The code solves these equations using a
conservative high-order Godunov scheme with an optimized approximate
Riemann solver \citep{toro97}. The algorithm is second-order accurate
in both space and time for smooth flows, and it provides robust
treatment of shocks and discontinuities. Although the code is capable
of solving the Poisson equation for the gravitational field $\phi$ of
the gas based on the
density distribution, in this work we neglect the gas self-gravity and
include only the gravitational force of the central object. The
potential is therefore
\begin{equation}
\phi = \frac{G M}{r},
\end{equation}
where $M$ is the mass of the central object and $r$ is the distance
from it. We do not adopt the \citet{paczynski80} potential frequently
used for simulations involving black holes because we are do not wish
to limit ourselves to the case of black holes, and because even in
the black hole case we are interested in the regime where the
Schwarzschild radius is extremely small on scales of the grid, so
that general relativistic effects are not important.

Our code operates within the AMR framework \citep{berger84,
berger89, bell94}, and is described in detail in \citet{truelove98} and 
\citet{klein99}. We discretize the problem domain onto a base, coarse
level, denoted level 0. We dynamically create finer levels, numbered
$1,2,\ldots n$, recursively nested within one another. To take a time
step, one advances level $0$ through a single time step $\Delta
t_0$, and then advances each subsequent level for the same amount of 
time. Each level has its own time step, and in general
$\Delta t_{l+1}< \Delta t_l$, so after advancing level 0 we must
advance level 1 through several steps of size $\Delta t_1$, until it
has advanced a total time $\Delta t_0$ as well. In all the simulations 
we present in this paper, we chose cell spacings such that $\Delta t_l 
= 2\Delta t_{l+1}$, and thus we take two time steps on level 1 for each
time step on level 0. (We find that refining by factors of two gives
better accuracy in the solution than using a larger refinement
factor.) After each level 0 time step, we apply
a synchronization procedure to guarantee conservation of mass,
momentum, and energy across the boundary between levels 0 and 1.
However, each time we advance level 1 through time $\Delta t_1$, we
must advance level 2 through two steps of size $\Delta t_2$, and
so forth to the finest level present.

On the finest level of
refinement, we represent the central, accreting object with a sink
particle \citep{krumholz04}. One noteworthy feature of our sink
particle, which we shall discuss further in \S\ref{sinkmethod}, is
that the sink particle does not accrete any angular momentum from the
gas in the computational grid. This is in contrast the sink
methodologies used by previous authors (e.g. \citet{ruffert97,
proga03}), where the sink could accrete angular momentum as well as
mass. Our approach is appropriate for $\oms\gg \omega_{\rm crit}$,
which is the case on which we focus.

\subsection{Initial and Boundary Conditions}

For each run, we place a sink particle at the origin of an initially
uniform gas. The gas has an initial velocity field
\begin{equation}
\mathbf{v_0} = \oms c_s \frac{y}{\rb} \hat{x},
\end{equation}
where we use different values of $\oms$ in different runs. Table
\ref{simtable} summarizes the values of $\oms$ that we
simulated. Our computational  domain extends from $-100\rb$ to
$100\rb$ in the $x$ and $z$
directions. In the $y$ direction, we also used a range of $-100\rb$ to
$100\rb$ for smaller values of $\oms$. For larger values of
$\oms$, though, the large velocities that occur at large values of
$y$ produce very small time steps that make the computation
prohibitively expensive. We therefore use a smaller domain in the
$y$ direction, as we indicate in Table \ref{simtable}; in every case,
however, we chose our domain to extend to values of $y$ such that
$v_0(y_{\rm max}) \gg v_{\rm escape}(y_{\rm max})$. 

We used inflow/outflow boundary conditions in the $x$ direction and
symmetry in the $y$
and $z$ directions. However, for all our runs the boundary is
sufficiently far away from the central object that, within the duration 
of the run, no sound waves can propagate from the central object to
the boundary and back. Our lowest resolution runs have $\Delta x_{\rm
min} < \rb/40$, which we use for smaller values of $\oms$.
For larger values of $\oms$, we use $\Delta x_{\rm
min} < \rb/160$. We set up our adaptive grids such that, at any radius 
$r$, the local grid resolution is $\Delta x\le \max(r/20,\Delta x_{\rm
min})$.

\subsection{The Sink Particle Method and Very Small Vorticity}

\label{sinkmethod}

As Table \ref{simtable} indicates, we vary $\oms$ from $10^{-2}$ to
$10^{1.5}$, 
thereby thoroughly exploring the small $\oms$ and large
$\oms$ regimes. However, none of our work explores the very
small $\oms$ regime. This is for two reasons, one physical and one
technical. In order to explore the very small $\oms$ case, one must
use either an 
unphysically small value of $\oms$ or an unphysically large accretor 
radius, leading to an unrealistically large $\omega_{\rm crit}$. As we
have shown in \S\ref{verysmall}, for typical astrophysical situations
in which accretion with vorticity is important, $\omega_{\rm crit}
\sim 10^{-3}$. This is a truly tiny amount of vorticity,
corresponding to a shear of one thousandth of the sound speed at one
Bondi radius. In the case of a protostar in a molecular clump, this
corresponds to a velocity gradient of no more than $\sim 10$ cm
s$^{-1}$ over a distance of $\sim 0.1$ pc; for the galactic center, it
corresponds to a gradient of $\sim 10^4$ cm s$^{-1}$ over a distance
of $\sim 0.1$ pc. It is difficult to see how to produce such an
irrotational flow field, and thus the very small $\oms$ case is not
relevant for the applications with which we are concerned.

The technical reason we do not treat the very small $\oms$ case is
that our sink particle method is constructed to exactly conserve total 
angular momentum during the accretion process. (For details, see
\citet{krumholz04}.) Transfer of vorticity from the flow to the accretor
is the distinguishing characteristic of flow with $\oms<\omega_{\rm
crit}$. However, since our sink particle does not change the angular
momentum of the flow field, it actually increases vorticity by
removing mass while leaving angular momentum. Our code does dissipate
vorticity via numerical viscosity, as we discuss in more detail in
\S\ref{convergence} on convergence testing. However, this effect is
small except in the inner few zones. On balance, we consider this
approach more realistic than the standard technique of allowing any
angular momentum that enters a chosen accretion region to accrete. In
reality, a small accretor should absorb negligible angular momentum
from the flow on scales much larger than the accretor radius. In real
accretion disks, viscosity acts to transfer mass inwards and angular
momentum outwards. This is exactly the approximation we adopt in our sink
particle method: mass travels inwards, angular momentum does not. 
While ideally one would follow the flow down to the true physical
surface of the accretor, this is computationally infeasible in more
than one dimension even with
AMR. Our sink particle method provides a more realistic approximation of
the behavior of an accretion disk than would allowing angular momentum 
to accrete.

\section{Simulation Results}

\label{simulations}

\subsection{Density and Velocity Fields}

\begin{figure}
\plotone{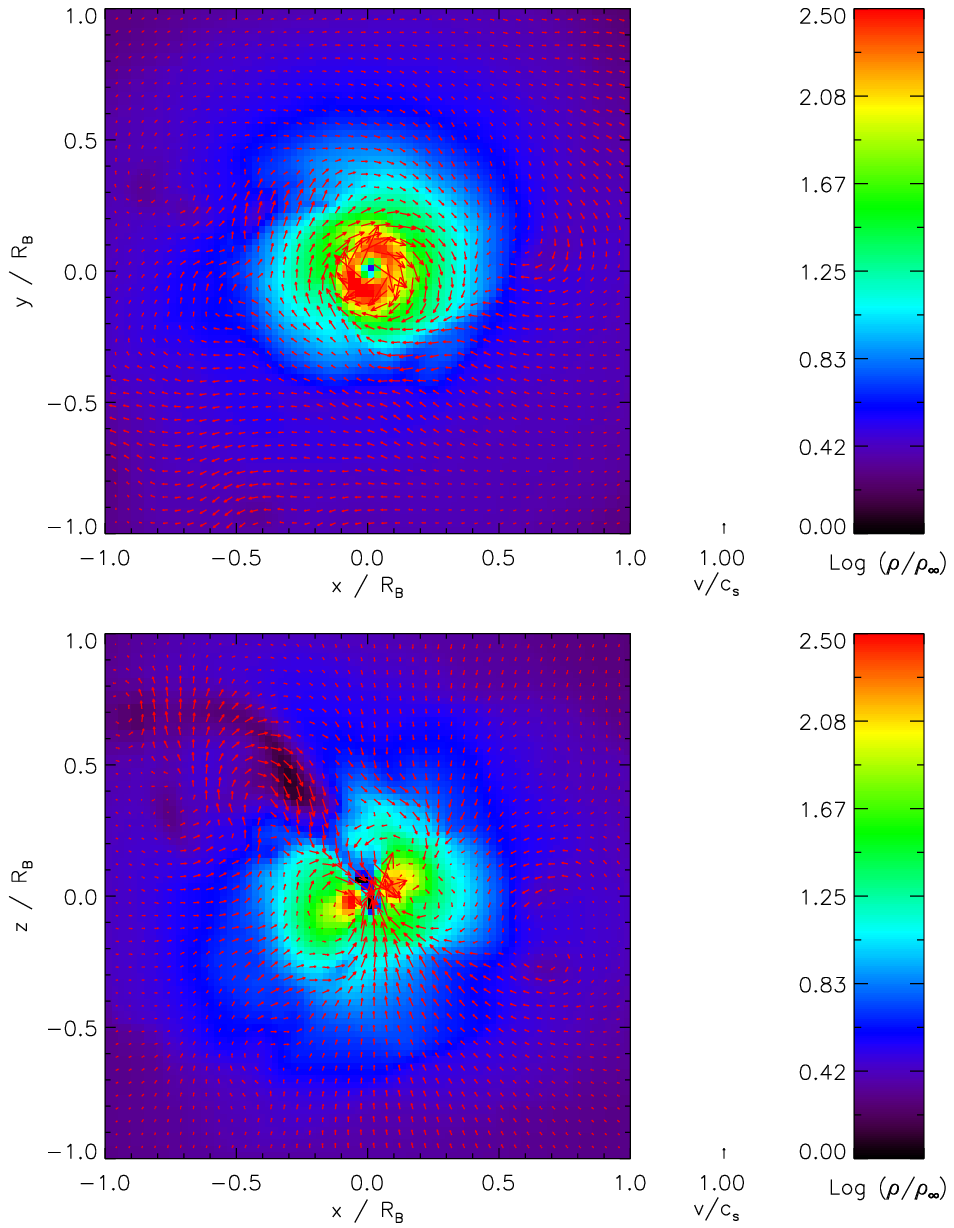}
\caption{\label{omloslice}
The plots show slices through the origin in the $xy$ and $xz$ planes
for the $\oms=10^{-2}$ run. The grayscale indicates log density, and 
the arrows indicate velocity. Note the differences in scale from
Figures \ref{ommedslice} and \ref{omhislice}.
}
\end{figure}

\begin{figure}
\plotone{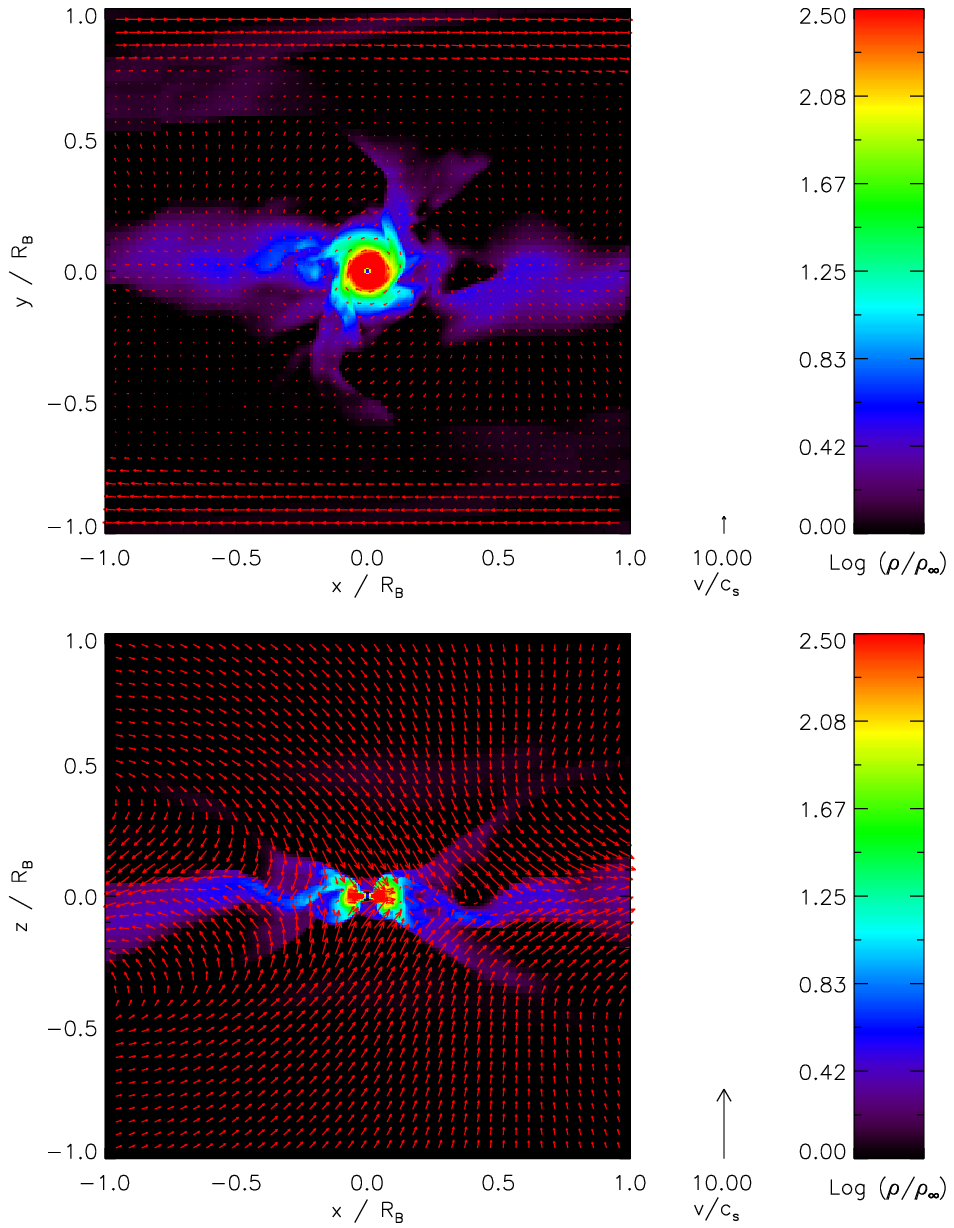}
\caption{\label{ommedslice}
The plots show slices through the origin in the $xy$ and $xz$ planes
for the $\oms=1$ run. The grayscale indicates log density, and 
the arrows indicate velocity. Note the differences in scale from
Figures \ref{omloslice} and \ref{omhislice}.
}
\end{figure}

\begin{figure}
\plotone{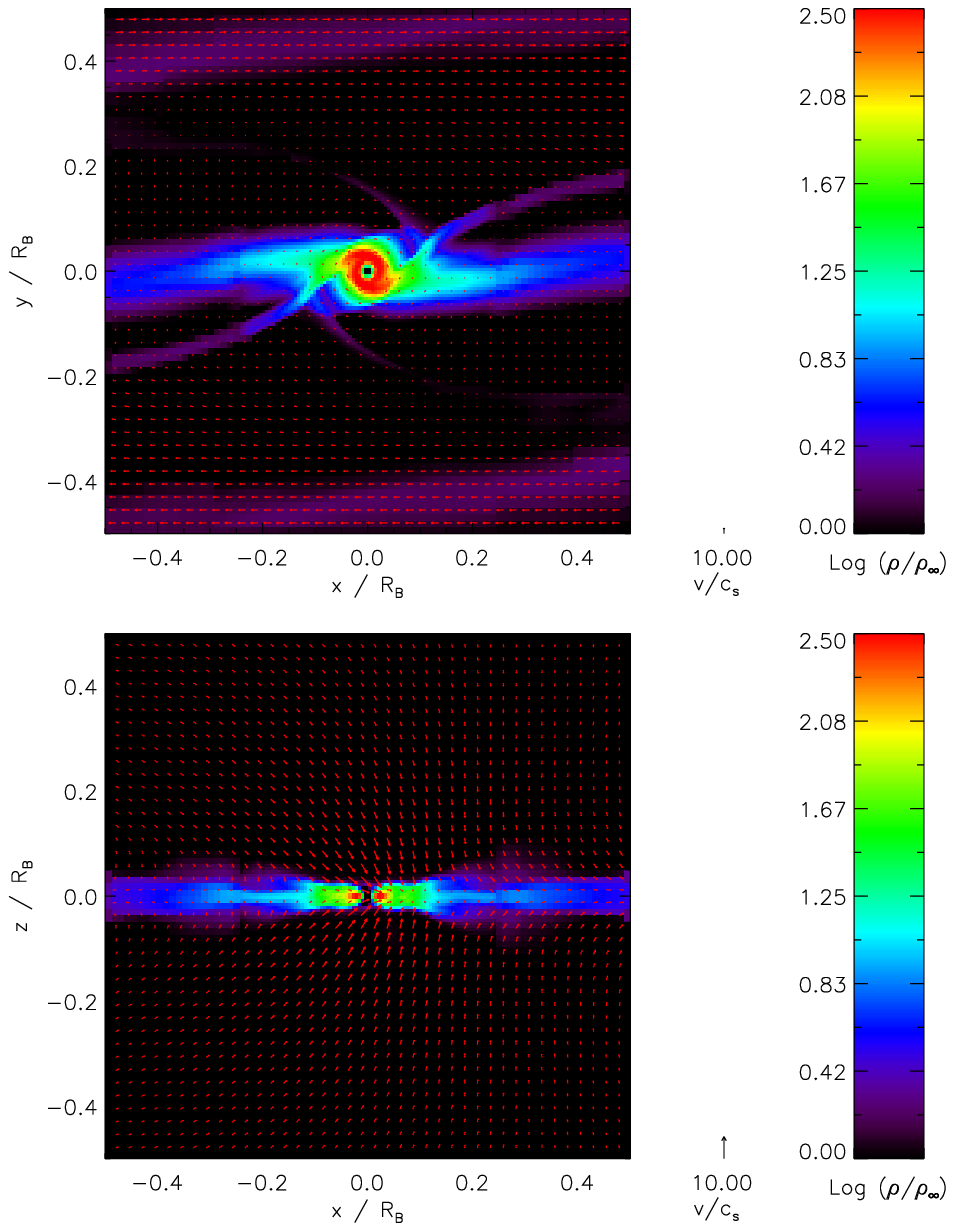}
\caption{\label{omhislice}
The plots show slices through the origin in the $xy$ and $xz$ planes
for the $\oms=10^{1.5}$ run. The grayscale indicates log density, and 
the arrows indicate velocity. Note the differences in scale from
Figures \ref{omloslice} and \ref{ommedslice}.
}
\end{figure}

In each of the simulations, the flow went through a transient and then
settled down into a quasi-equilibrium state. The time required to
reach equilibrium increased with decreasing $\oms$, ranging from
$\sim 100\tb$ for $\oms=10^{-2}$ to $\sim \tb$ for $\oms\gtsim 1$, 
where $\tb\equiv \rb/c_s$ is the Bondi time. In the equilibrium
configuration, all the runs except $\oms=0$ built up dense material 
around the accretor. The material was not in a symmetric disk, but
rather in a pinwheel shape. The velocity patterns associated with
these pinwheels generally involved matter flowing in from the $z$
direction through an accretion funnel and either circulating or
outflowing in the $xy$ plane. This overall morphology agrees well with
the results of  \citet{proga03}. Figures \ref{omloslice},
\ref{ommedslice}, and \ref{omhislice} illustrate the morphology for
$\oms=10^{-2}$, 1, and $10^{1.5}$.

A comparison of runs at different $\oms$ values shows several clear
trends. First, the accumulated pinwheel of matter is much smaller
at large $\oms$ than at small $\oms$. For $\oms
\ltsim 1$, the accumulated matter extends out to $\sim\rb/2$, while
for $\oms\gg 1$ it is well inside the Bondi radius and is less
massive. The size of the pinwheel changes only slightly with $\oms$
for $\oms\ltsim 1$, but shrinks dramatically for $\oms\gtsim 1$. Our
theoretical description of the torus we expect for small $\oms$
appears to be roughly correct: a torus of material accumulates that
extends out to $\sim\rb$ and has a height comparable to its
radius. The tori are chaotic and time-variable. For larger $\oms$,
the pinwheels are more regular and less chaotic, and are accompanied by
leading and trailing shocks. The shocks are moderately strong, with
Mach numbers $\sim 2-3$.

For $\oms\ltsim 1$, we do not find that the torus has constant
specific angular momentum, as \citet{proga03} did. In our torii, the
specific angular momentum of the gas varies by orders of magnitude.
It is unclear if this difference in results is due to
different initial conditions, or due to the fact that our simulations
are three-dimensional where theirs were two-dimensional. However, we
agree with \citet{proga03} that the size and shape of the torus is
largely independent of $\oms$ (or the initial specific angular
momentum $l$), and that its equilibrium structure arises due to the
accumulation of material with too much vorticity (or $l$) to accrete.
As Figures \ref{omloslice} and \ref{ommedslice} show, accretion
occurs only through a narrow funnel, and the (invariant) shape of the
torus therefore determines the accretion rate.

\subsection{Evolution of Circulation}

For each run we computed the average dimensionless circulation $\Gamma$
as a function of time. We define the dimensionless circulation within
a circle of radius $\rb$ in the $xy$ plane as
\begin{equation}
\Gamma_{*} = -\frac{1}{\pi\rb c_s}
\oint_{r=\rb} \mathbf{v}\cdot \mathbf{dl},
\end{equation}
where the line integral is evaluated on the circle $r=\rb$. The
circulation is a measure of the amount of vorticity within one Bondi
radius of the accretor. With this definition of dimensionless
circulation, in the initial configuration of the velocity field the
circulation is equal to the vorticity parameter, $\Gamma_*=\oms$. Our
prediction for the equilibrium circulation for $\oms\ltsim 1$ is
$\Gamma\sim 2\pi c_s \rb$, or $\Gamma_*\sim 2$.

\begin{figure}
\plotone{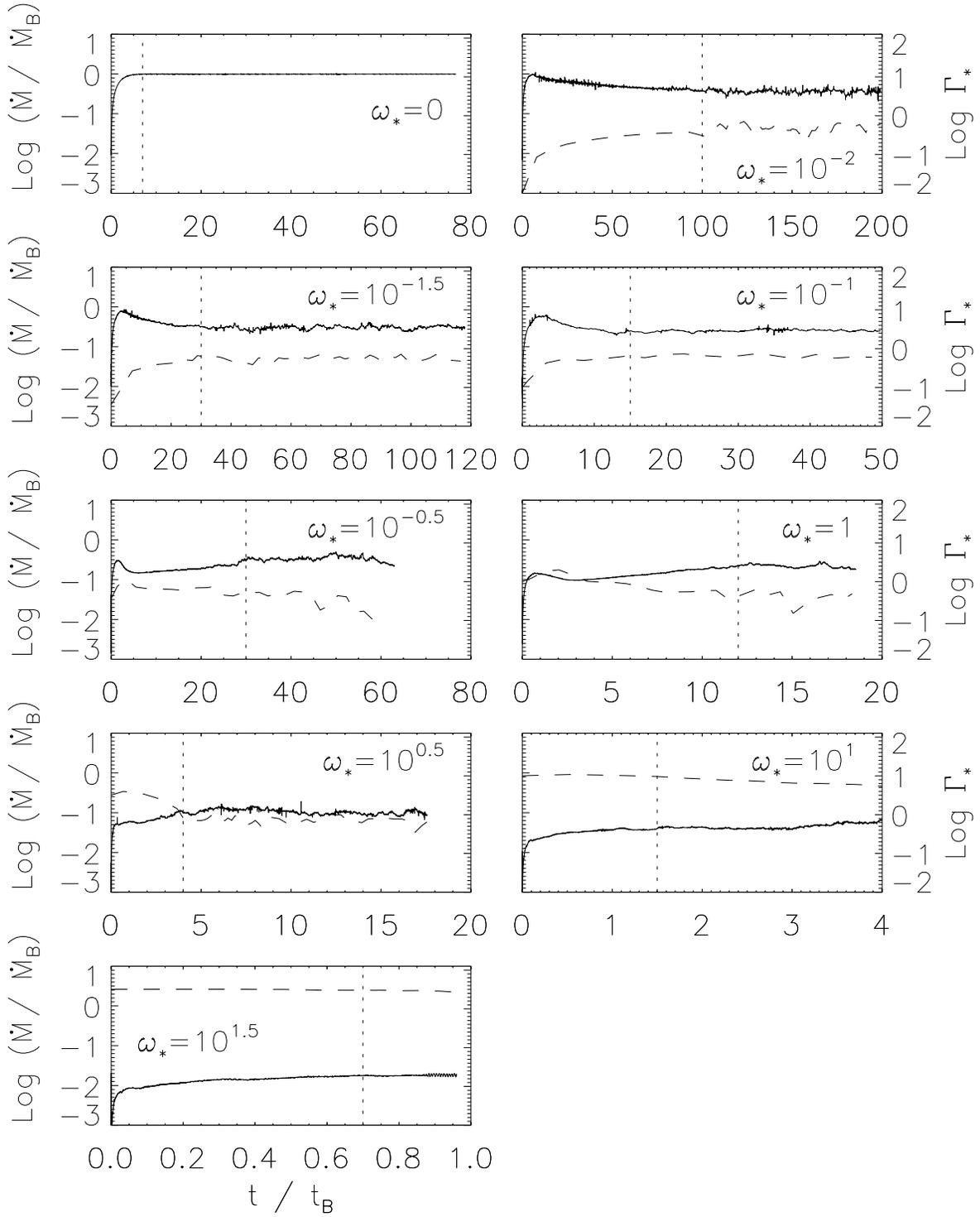}
\caption{\label{accpanel}
The plots show the accretion rate versus time (solid lines) and
circulation versus time (dashed lines) for the run with the value of
$\oms$ indicated in the panel. The dotted vertical lines indicate the
time at which we conclude the system has reached equilibrium.
}
\end{figure}

We plot the circulation versus time along with the accretion rate
versus time for each of our runs in Figure \ref{accpanel}.
The behavior of the circulation with time depends on
the value of $\oms$. For $\oms\ltsim 1$, the circulation starts at
$\oms$ and gradually increases until it reaches $\sim 0.4-0.8$. Then
it fluctuates around this mean. For $\oms=10^{0.5}$,
$\Gamma_*$ decreases from its initial value until it again reaches the
range
$\sim 0.4-0.8$. As $\oms$ increases, however, $\Gamma_*$ decreases
less and less in the time it takes the accretion rate to reach
equilibrium. The accretion rate and the circulation appear to be
anti-correlated for $\oms \ltsim 1$ in the period before
equilibration: the initial buildup of circulation takes about the same
amount of time as the initial decrease in the accretion rate. This
supports our hypothesis that vortex lines are acting like magnetic
flux lines and inhibiting accretion.

\begin{figure}
\plotone{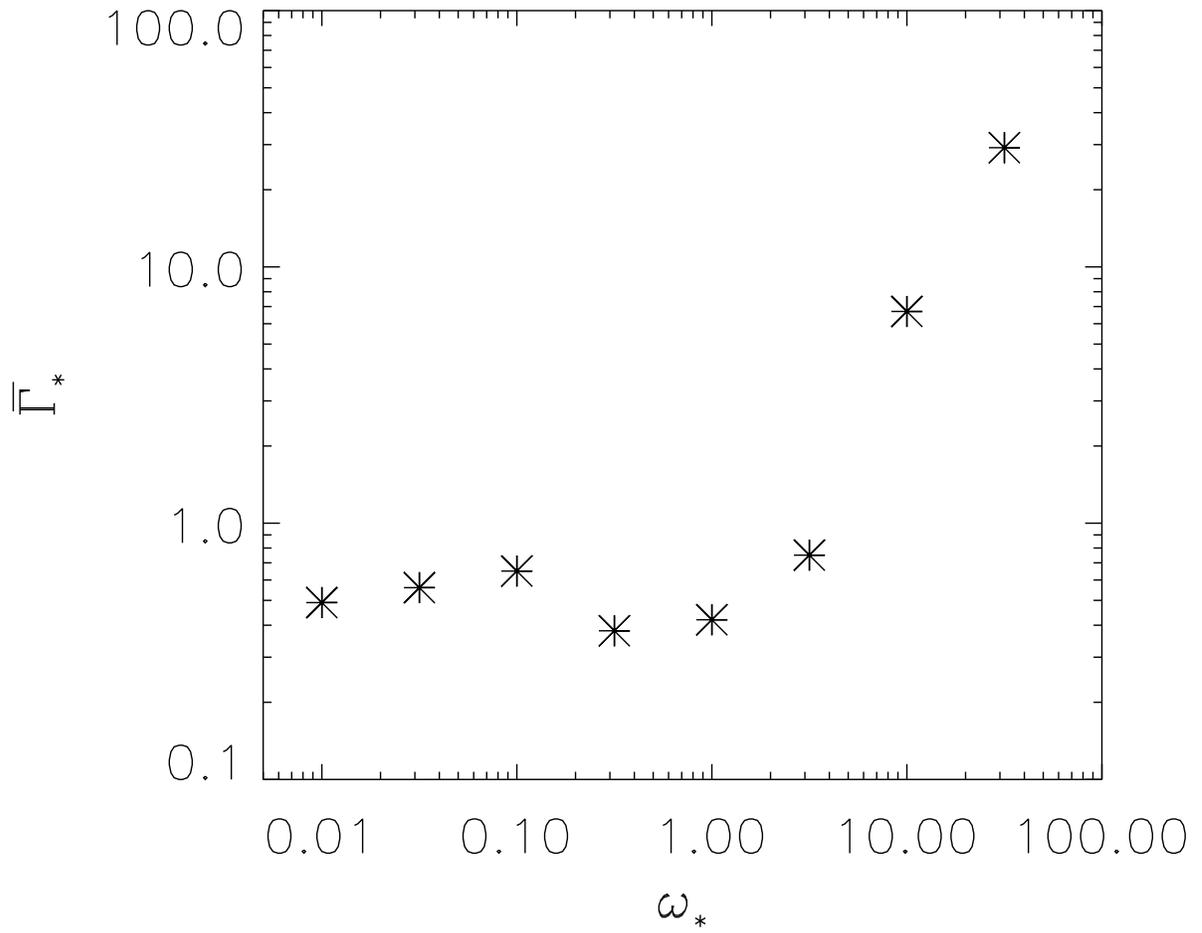}
\caption{\label{gammaplot}
The asterisks indicate the mean value of $\Gamma_*$ after equilibration
for each of our runs as a function of $\oms$.
}
\end{figure}

\begin{figure}
\plotone{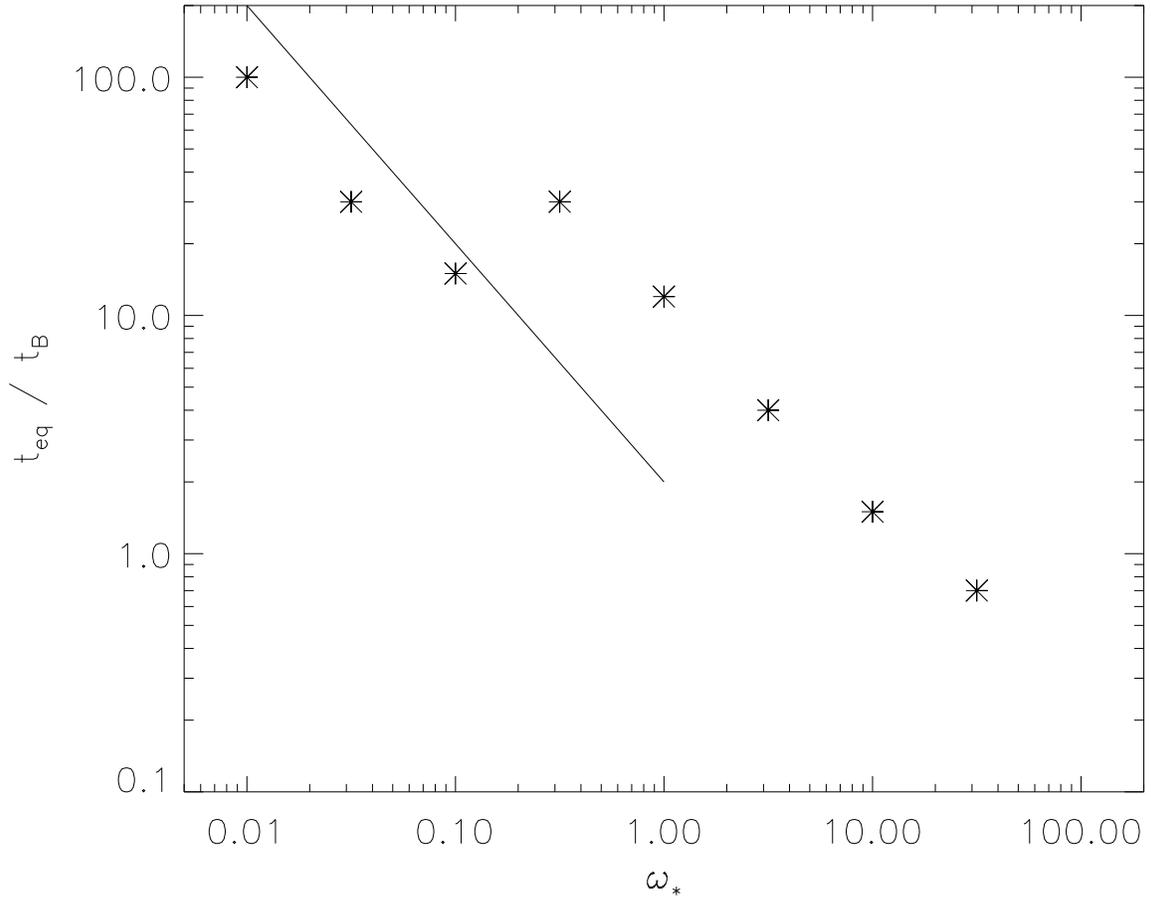}
\caption{\label{teqplot}
The asterisks indicate the time (in units of $\tb$) at which the run
reaches equilibrium for each value of $\oms$. The line is our
prediction $t_{\rm eq}=2\tb/\oms$.
}
\end{figure}

For each run we estimate by eye the time at which the flow pattern and
accretion rate reach equilibrium, and we compute the mean circulation
after this time. We report the equilibration times and the mean
circulations in Table \ref{simtable}; we plot
the equilibrium value of $\Gamma_*$ in Figure \ref{gammaplot}, and
the equilibration time in Figure
\ref{teqplot}. As the figures show, the
equilibrium value of the circulation is at most very weakly dependent
on the initial vorticity for $\oms\ltsim 1$, varying by less than a
factor of 2 while the initial vorticity varies by a factor of
100. The flow pattern re-arranges itself to select $\Gamma\sim
0.4-0.8$ by accumulating a dense torus of material rotating at nearly 
Keplerian speeds around the accretor. The equilibrium value is
smaller than our prediction of $\Gamma_*\sim 2$, indicating
that the rotational velocity is sub-Keplerian. This result is
consistent with the morphology of the gas flow. The gas has a non-zero
radial velocity and thus can be marginally bound even with tangential
velocities that are sub-Keplerian. Our analytic model, because it
fails to predict the chaotic nature of the flow pattern,
overestimates the equilibrium value of $\Gamma_*$ by a factor of $\sim 
4$. It does, however, correctly predict that there is an equilibrium
$\Gamma_*$ for $\oms\ltsim 1$. At $\oms\gtsim 1$, the pinwheel
of unaccreted matter is confined closer to the accretor, and is unable 
to affect the circulation on length scales $\sim\rb$. Thus, the
circulation for larger $\oms$ stays constant.

The equilibration time is also roughly consistent with our prediction
of $t_{\rm eq}=2 \tb/\oms$ for $\oms\ltsim 1$. Part of the scatter of
the line comes from the unavoidable subjectivity in our estimate of
the equilibration time in such a chaotic flow. However, our model is
clearly only approximate. While we correctly capture the trend of
increasing equilibration time with decreasing $\oms$, our inability to 
model the turbulent flow means that our prediction of the
equilibration time, like our prediction of $\Gamma_*$, is accurate to
at best a factor of a few.

In the plots of $\Gamma_*$ and $t_{\rm eq}$ versus $\oms$ (and as 
we show below, in the plot of $\mdot$ versus $\oms$ as well), there
seems to be a change in behavior between the $\oms=10^{-1}$ and
$\oms=10^{-0.5}$ runs. Rather than smoothly interpolating bewteen
$\oms\ll 1$ and $\oms\gg 1$, the trend with $\oms$ appears to jump
from one track to another. We cannot rule out the possibility that
this is simply the result of chance and our sparse sampling $\oms$
values. Near $\oms=1$, the equilibrium values of quantities like
$\Gamma_*$ and $t_{\rm eq}$ may be fluctuating chaotically with
$\oms$, and the sharp jump we see from $\oms=10^{-1}$ to
$\oms=10^{-1/2}$ may just be a fluctuation. However, it is also
possible that there is a distinct regime of intermediate $\oms$, as
well as the small and large $\oms$, and that several properties of the
equilibrium flow change suddenly when one enters this new regime.

\subsection{Accretion Rates and Comparison to Theory}

We ran each of our simulations until the accretion rate converged to a 
steady state, and then we measured both the mean accretion rate and the
fluctuations about the mean. Table \ref{simtable} summarizes our
results and Figure \ref{accpanel} shows the accretion rate versus time
for each run. We filter out short time
scale noise associated with the finite resolution of our simulation by
only computing the standard deviation in accretion rate averaged over
16 time
steps. Nonetheless, discreteness error does lead to a non-zero
standard deviation even in the case $\oms=0$, which is pure Bondi
accretion and should have zero standard deviation. The standard
deviation of $0.007\%$ gives a lower limit on our ability to measure
fluctuations in the accretion rate. Since the standard
deviations we measure for other runs are significantly larger than
this, they likely represent real physical time variability, not just
simulation noise.

\begin{figure}
\plotone{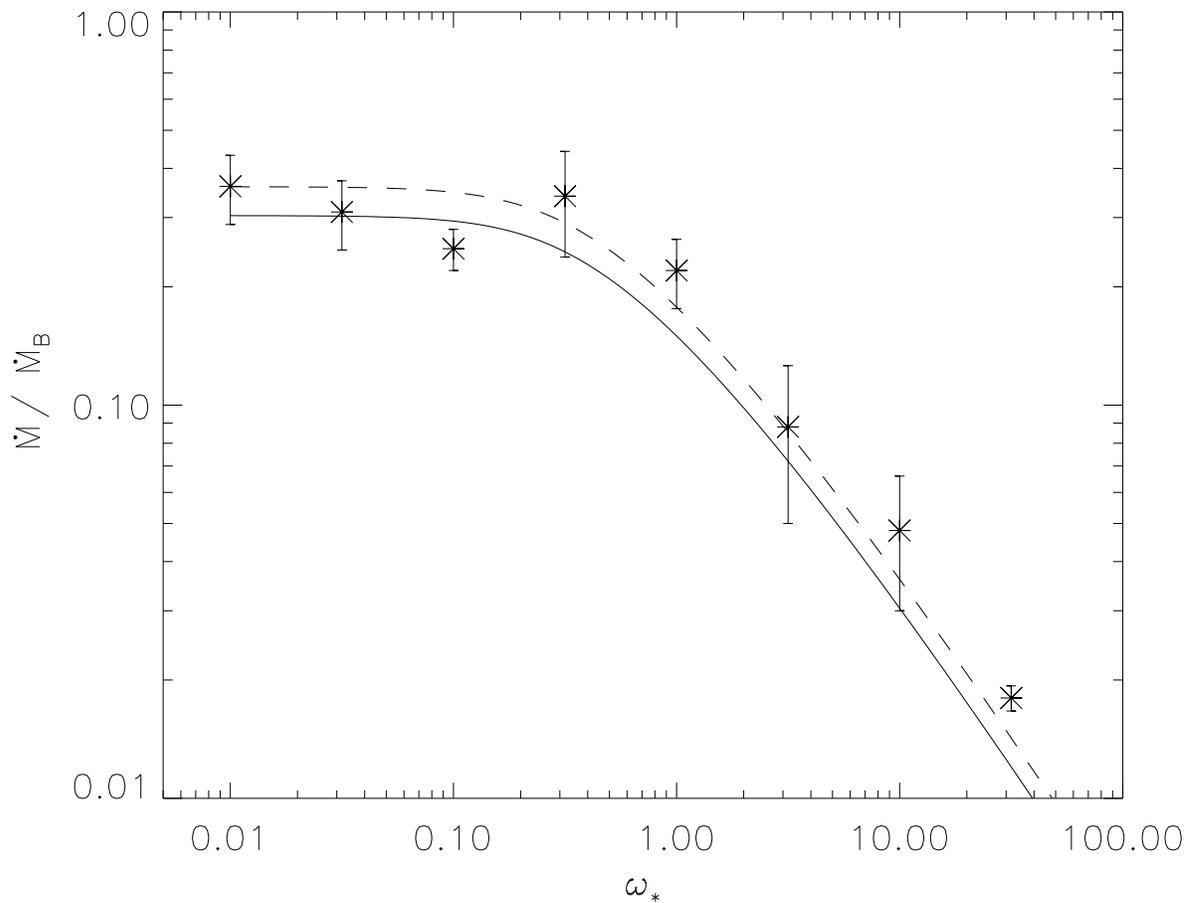}
\caption{\label{accplot}
The plot shows mean accretion rate, in units of the Bondi rate, versus
$\oms$. The asterisks are the simulation data, with error bars
indicating two times the standard deviation in the accretion rate. The 
solid line is our theoretical prediction (\ref{mdotapprox}). The
dashed line is the theoretical prediction scaled by a constant factor
to give the best possible fit to the simulation data.
}
\end{figure}

We plot the mean accretion rate as a function of $\oms$ in Figure
\ref{accplot}. We also show our theoretical prediction (solid line),
which fits the data reasonably well. To obtain a somewhat better fit,
we keep the function $f(\oms)$ the same and do a least squares fit
to obtain a pre-factor to replace 0.34 in equation
(\ref{mdotapprox}). We find a best-fit value of 0.40. This value
produces a good fit, as shown in the dashed line in Figure
\ref{accplot}. The formula fits each of the simulation values to
better than 40\%. This is comparable to the accuracy of the
Bondi-Hoyle formula at intermediate Mach numbers
\citep{ruffert94a,ruffert94b}.

\subsection{Convergence Testing and Dependence on $\ra/\rb$}

\label{convergence}

To test the convergence of our simulations, we repeated one of our
runs at higher resolution. \citet{perna03} have hypothesized that the
accretion rate in Bondi accretion with vorticity is proportional to
$(\ra/\rb)^p$, where $p$ is between $0.5$ and $1$. \citet{proga03}
saw a small dependence of the accretion rate on $\ra/\rb$ for some of
their simulations with smaller dynamic ranges, but the variation
seemed to disappear in their simulations with larger dynamic
ranges. In our sink particle method, the central object cannot accrete
angular momentum, so in the absence of numerical viscosity, we would
be in the limit $\ra/\rb\ll 1$. However, as we have noted
above, numerical viscosity does set an effective minimum size for our
accretor. In the inner few cells of the simulation grid, where the
circles in which the gas is attempting to move are poorly resolved,
our code does not perfectly conserve angular momentum or
vorticity. It is difficult to determine an equivalent accretor size
set by this phenomenon, but we have found that even at our lowest
resolution of $\rb/\Delta x=40$, our simulations produce 
disks and accumulation of circulation rather than Bondi-like flow for
$\oms$ as small as $10^{-3}$. Thus, the effective value of $\ra/\rb$ set
by our method must be $\ltsim 10^{-3}$. (Recall that typical
values of $\ra/\rb$ in astrophysical systems are closer to $10^{-6}$.)
Regardless of the true effective accretor radius imposed by numerical
viscosity, by changing the grid size, and thus the rate of angular
momentum dissipation, we are testing whether the accretion rate truly
depends on $\ra/\rb$ when $\ra/\rb\ll 1$.

\begin{figure}
\plotone{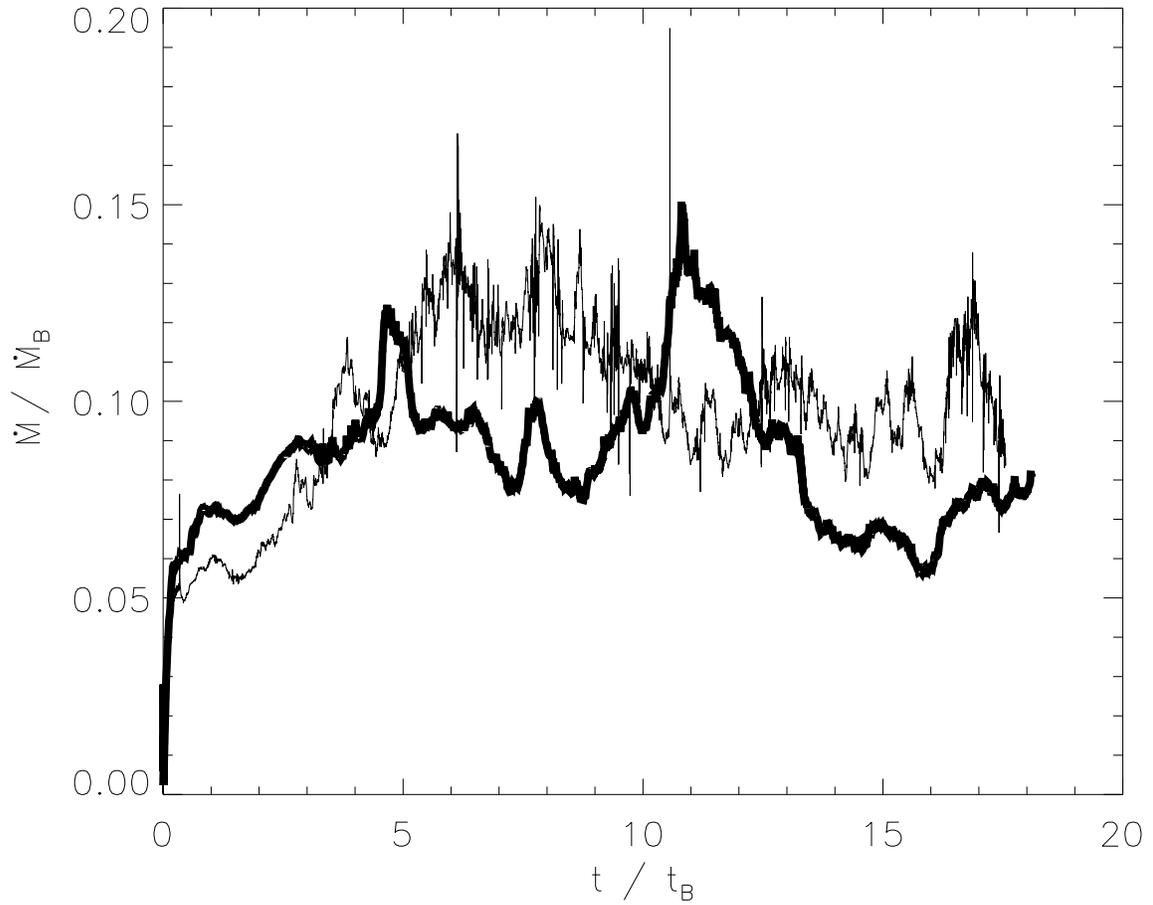}
\caption{\label{convfig}
The plot shows the accretion rate versus time for our convergence
test. The thick line is the $\oms=10^{0.5}$ run at a resolution of
$\rb/\Delta x = 40$, and the thin line is the run at a resolution of
$\rb/\Delta x = 160$.
}
\end{figure}

To test for convergence, we re-ran our simulation with $\oms=10^{0.5}$
at a resolution of $\rb/\Delta x=160$, four times the fiducial
resolution. All other aspects of the
simulation were identical. We plot the accretion rate 
versus time for the two differently resolved versions of the run in
Figure \ref{convfig}. As the results show, the exact shape of the accretion
rate versus time graph is not the same at the different
resolutions. This is to be expected in an unstable and chaotic
flow. However, the overall accretion rates appear to be approximately
the same. We find a mean accretion rate after equilibration of
$0.088\mdotb$ for the run with $\rb/\Delta x=40$, and a mean of
$0.11\mdotb$ for the run with $\rb/\Delta x=160$. The standard
deviations in the accretion rates are $22\%$ and $14\%$, respectively, 
while the difference in means is only $20\%$. Thus, our results are
consistent with the hypothesis that there is no dependence of the
accretion rate on $\ra/\rb$ in the limit $\ra\ll \rb$. The results
also suggest that our simulations are converged. In contrast, consider 
the hypothesis of \citet{perna03} that the accretion rate scales as
$(\ra/\rb)^p$ for $p$ in the range 0.5 to 1. By increasing the
resolution by a factor of 4, we should have decreased $\ra/\rb$ by a
factor of 4, and thus the \citet{perna03} proposal predicts we should
have measured an accretion rate in the range $0.044\mdotb$ to
$0.022\mdotb$ for the higher resolution simulation. This is clearly
inconsistent with our results. We emphasize, however, that we have
only tested the hydrodynamic case. \citet{perna03} made their
hypothesis for both hydrodynamic and magnetohydrodynamic flows, and we 
have not tested the latter case.

\section{Conclusions}

\label{conclusions}

We provide a general theoretical framework for considering problems of 
accretion in a medium with vorticity. Using simple analytic estimates, 
we have been able to derive a formula for the rate of mass accretion
onto a point particle of radius $\ra$ as a function of the vorticity
present in the ambient medium. We define the dimensionless vorticity
parameter $\oms$ as
\begin{equation}
\oms = |\mbox{\boldmath$\omega$\unboldmath}| \frac{\rb}{c_s},
\end{equation}
where $\rb$ is the accretor's Bondi radius and $c_s$ is the sound speed.
We find that the accretion rate is
\begin{equation}
\mdot(\oms) \simeq
4\pi \rho_{\infty} \frac{(GM)^2}{c_s^3}\, 0.4\, f(\oms),
\end{equation}
for $\oms\gtsim\omega_{\rm crit}$, where $f(\oms)$ is
the function defined by (\ref{fomega}). We note that $f(\oms)$ has the 
limiting behavior $f(\oms)\approx 1$ for $\oms\ll 1$, and
\begin{equation}
f(\oms) \approx \frac{2}{3\pi\oms} \ln\left(16\oms\right)
\end{equation}
for $\oms\gg 1$. Simulations show that our formulation provides a
good fit to the overall shape of the curve of accretion rate versus
vorticity. By calibrating our first-principle calculation using the
results of our simulations, we provide a formula that agrees with the
mean accretion rate we measure in our simulations to better than 40\%,
all over 3.5 orders of magnitude variation in vorticity. Our result is
a natural extension of the Bondi-Hoyle-Lyttleton formula to rotating
flows.

We are also able to roughly predict several other properties of the
flow, including the overall morphology, the equilibrium vorticity, and
the equilibration time. Our predictions for these quantities are
considerably less accurate than for the accretion rate, but our
formulae are correct to within factors of a few. We see ambiguous
evidence for the existence of an intermediate $\oms$ regime, which is
characterized by a rapid change in equilibrium flow properties
starting between $\oms=0.1$ and $\oms=10^{0.5}$. Our analytic model
does not predict such the existence of such 
a transition, and it is possible that what we have seen is a result of 
numerical noise. Regardless, it does seem clear that the analytic
model runs into trouble when $\oms\sim 1$. Since we derived it for the 
regimes $\oms\ll 1$ and $\oms \gg 1$, and simply interpolated between
those, this is not surprising.

The simulations we use to check our formulation have very general
and simple initial conditions. We have avoided complications arising
from boundary conditions by moving the boundaries very far away from
the accreting object, and our sink particle method allows us to avoid
the need for an unrealistically large accretor radius. Our theoretical 
discussion shows that the accretion rate is controlled by a
combination of vorticity build-up near the accretor and, for
$\oms\gtsim 1$, the rate at which gas with low enough energy to
accrete approaches the accreting particle. As a result, our results
should be extremely robust.

Our results allow direct application to observed systems. One
may determine the vorticity in a astrophysical system by
observing a velocity gradient across it, which is often
feasible using radio line observations. Consider an
example in the context of star formation:
\citet{goodman93} observed dense molecular cores in nearby star-forming
regions. They found velocity gradients of $\sim 0.3-3$ km s$^{-1}$
pc$^{-1}$, in regions with typical temperatures of $\sim 10$ K (sound
speed $c_s\approx 0.2$ km s$^{-1}$ for a mean molecular mass of
$2.33m_{\rm H}$). This means that a one solar mass
protostar ($\rb\approx 0.1$ pc) inside such a dense core has a
vorticity parameter $\oms \sim 0.15-1.5$ (specific angular momentum
$l\sim 8\times 10^20-8\times 10^{21}$ gr cm s$^{-1}$, and should have its
accretion rate reduced by a factor of $\sim 3-7$ relative to the Bondi 
rate. This result has potential implications for models of star
formation in which protostars gain mass through a process of
competitive accretion.

\acknowledgments We thank Eliot Quataert for useful discussions, and
the referee for helpful comments. This
work was supported by: NASA GSRP grant NGT 2-52278 (MRK); NSF grant
AST-0098365 (CFM); NASA ATP grant NAG 5-12042 (CFM and RIK);
and the US
Department of Energy at the Lawrence Livermore National Laboratory
under contract W-7405-Eng-48 (RIK). This research used resources of
the National Energy Research Scientific Computing Center, which is
supported by the Office of Science of the U.S. Department of Energy
under Contract No. DE-AC03-76SF00098, through ERCAP grant 80325, 
and the NSF San Diego
Supercomputer Center through NPACI program grant UCB267.

\appendix

\section{Approximation of $f(\oms)$ for Large $\oms$}

\label{integralapprox}

\begin{figure}
\plotone{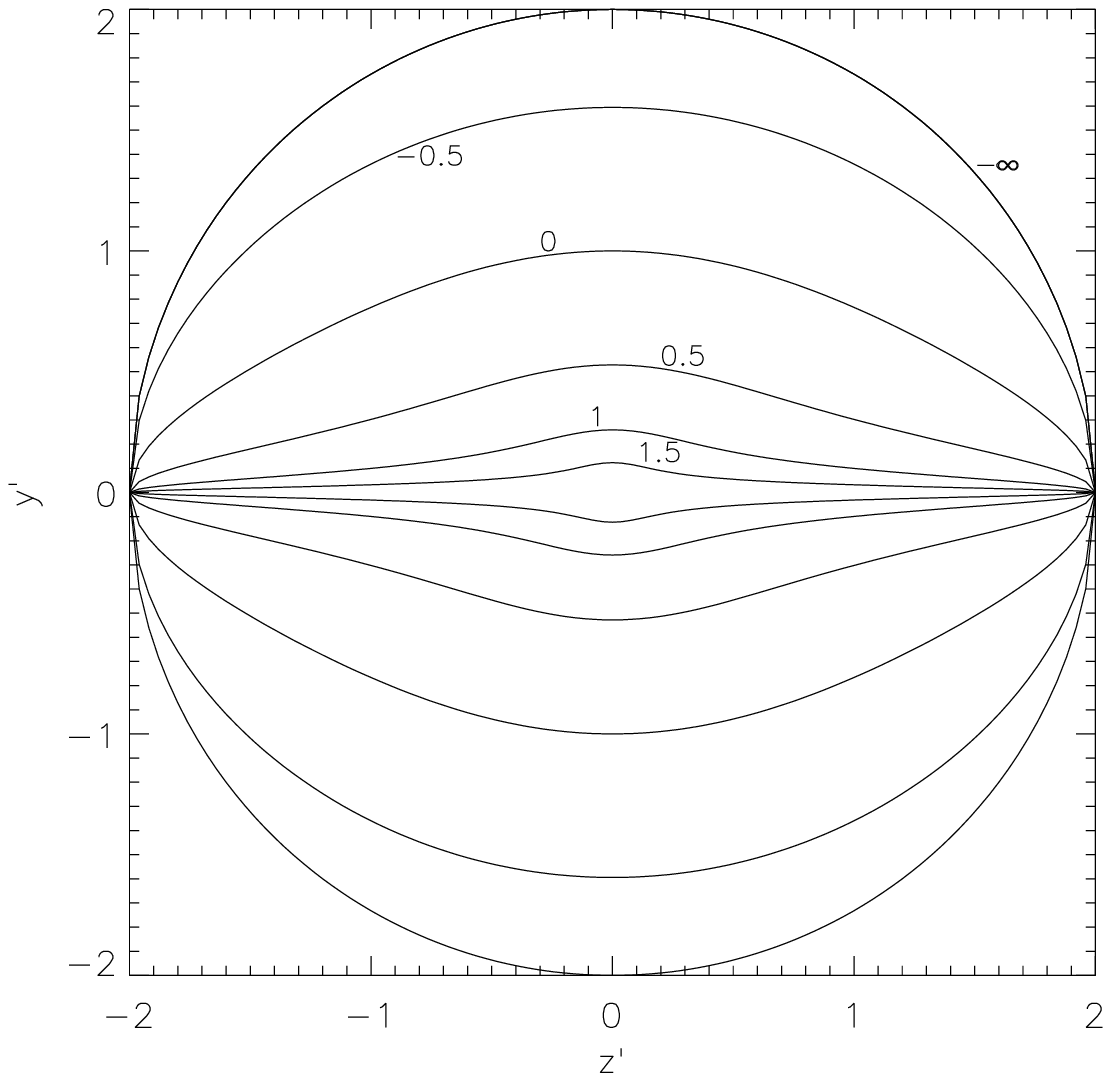}
\caption{\label{Aplot}
The plot shows the area $A(\oms)$, as defined by equation
(\ref{Aomega}), for different values of $\oms$. The numbers indicate 
the value of $\log\oms$ for each curve.
}
\end{figure}

We wish to derive an approximate formula for
\begin{equation}
f(\oms) = \frac{1}{\pi}\int_0^{y'_{\rm max}(\oms)}
	\sqrt{\frac{4-y'^2 (1+\oms^2 y'^2)^2}{1+\oms^2 y'^2}} dy'
\end{equation}
in the limit $\oms\rightarrow\infty$. We begin by breaking it into two 
terms,
\begin{equation}
f(\oms) = \frac{1}{\pi}\int_0^{y'_{\rm max}(\oms)}
	\sqrt{\frac{4}{1+\oms^2y'^2}-y'^2 (1+\oms^2 y'^2)} \; dy'.
\end{equation}
At the upper limit of the integration, $y'=y'_{\rm max}$, the
two terms under the square root are equal and the integrand
vanishes. For smaller values of $y'$, the first term goes to $4$,
while the second term scales as $y'^2$. Thus, the second term is
significant only near the upper limit of integration. Examining Figure 
\ref{Aplot}, we can see that the region $y'\approx y'_{\rm max}$ makes
only a small contribution to the integral in the large $\oms$
case. Thus, we may obtain an approximation for $f(\oms)$ when $\oms\gg
1$ by dropping the second term, giving
\begin{eqnarray}
f(\oms) & \approx & \frac{1}{\pi}\int_0^{y'_{\rm max}(\oms)}
	\sqrt{\frac{4}{1+\oms^2y'^2}} \; dy' \\
\label{fapprox1}
& = & \frac{2}{\pi \oms} \sinh^{-1}
	\left[\oms y'_{\rm max}(\oms)\right].
\end{eqnarray}
The upper limit of integration is
\begin{equation}
y'_{\rm max}(\oms) =
\frac{3^{1/3} \left(9\oms + \sqrt{3+81\oms^2}\right)^{2/3} - 3^{2/3}}
{3 \oms \left(9\oms+\sqrt{3+81\oms^2}\right)^{1/3}},
\end{equation}
which for $\oms\gg 1$ behaves as
\begin{equation}
\lim_{\oms\rightarrow\infty} y'_{\rm max}(\oms) =
\left(\frac{2}{\oms^2}\right)^{1/3}.
\end{equation}
Substituting into (\ref{fapprox1}), we find
\begin{equation}
f(\oms) \approx \frac{2}{\pi \oms} \sinh^{-1}
	\left[ \left(2 \oms\right)^{1/3} \right].
\end{equation}
Taking the limit $\oms\rightarrow\infty$, this becomes
\begin{equation}
f(\oms) \approx \frac{2}{3\pi\oms} \ln\left(16\oms\right).
\end{equation}
Numerical integration shows that this approximation is fairly accurate 
over a wide range of $\oms$, but the convergence is logarithmic and
the approximation does not become very good until $\oms$ is quite
large. It is within $10\%$ for $\oms>4$ and within $5\%$ for
$\oms> 53$, but does not become accurate to $1\%$ until $\oms>
1.8\times 10^{12}$.

\clearpage

\clearpage

\begin{figure}
\plotone{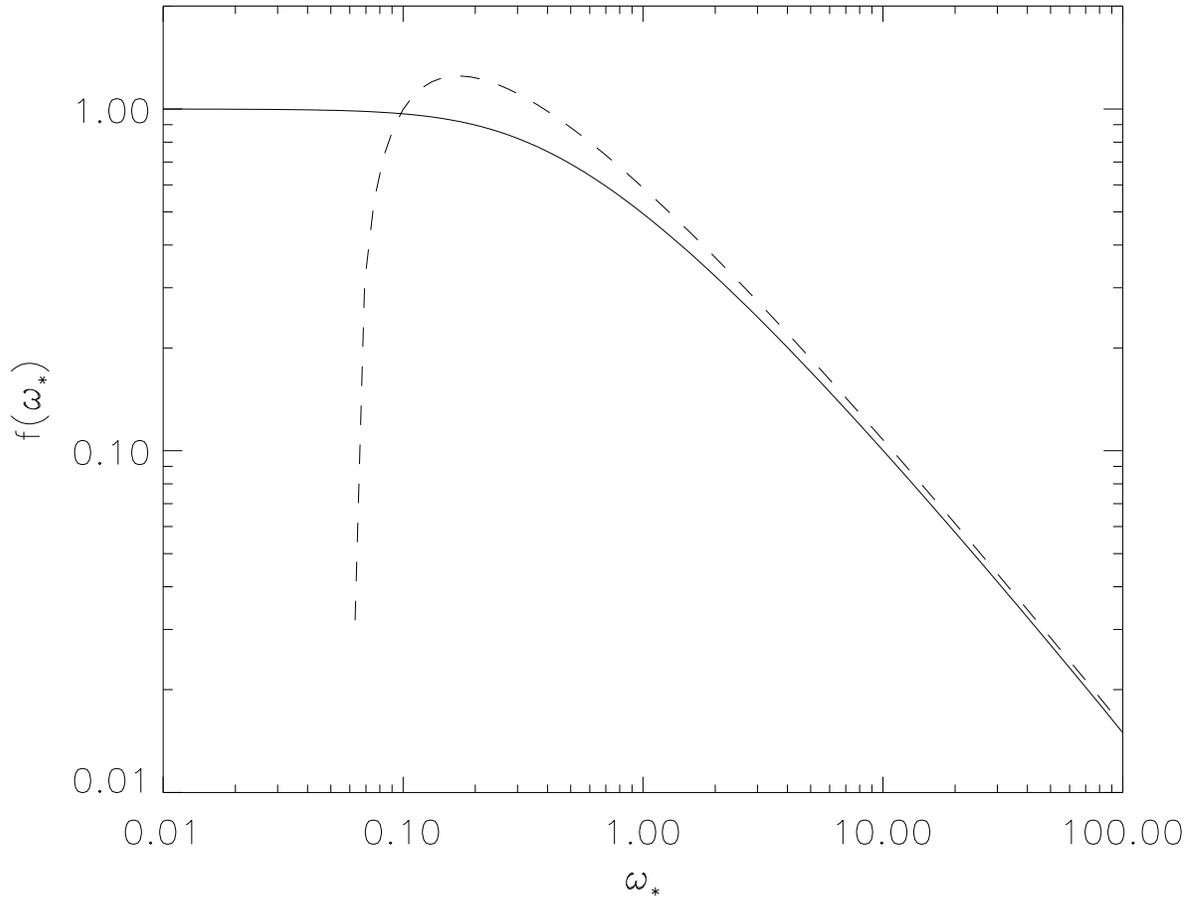}
\caption{\label{fplot}
The solid shows $f(\oms)$, as defined by equation (\ref{fomega}). The
dashed line shows the large $\oms$ approximation given by equation
(\ref{fapprox}).
}
\end{figure}

\clearpage

\begin{deluxetable}{cccccccc}
\tablecaption{Simulation parameters and results.\label{simtable}}
\tablewidth{0pt}
\tablehead{
\colhead{$\oms$} &
\colhead{$\Delta x_{\rm min}/\rb$} &
\colhead{$y_{\rm range}/\rb$} &
\colhead{$t_{\rm run}/\tb$} &
\colhead{$t_{\rm eq}/\tb$} &
\colhead{$\overline{\mdot}/\mdotb$} &
\colhead{$\sigma_{\dot{M}}$} &
\colhead{$\overline{\Gamma}_*$}
}
\startdata
$0$ 		& 0.025  & $[-40,40]$     & 77   & 7	& 0.99	& 0.007
& $2.3\times 10^{-4}$ \\
$10^{-2}$ 	& 0.025  & $[-40,40]$     & 200  & 100	& 0.36	& 0.099 
& 0.49 \\
$10^{-1.5}$ 	& 0.025  & $[-40,40]$     & 120  & 30	& 0.31	& 0.099
& 0.56 \\
$10^{-1}$ 	& 0.025  & $[-40,40]$     & 50   & 15	& 0.25	& 0.061
& 0.65 \\
$10^{-0.5}$ 	& 0.025  & $[-40,40]$     & 63   & 30	& 0.34	& 0.15
& 0.38 \\
$10^{0}$ 	& 0.025  & $[-40,40]$     & 19   & 12	& 0.22	& 0.10
& 0.42 \\
$10^{0.5}$ 	& 0.025  & $[-12.5,12.5]$ & 18   & 4	& 0.088	& 0.22
& 0.75 \\
$10^{1}$ 	& 0.0061 & $[-6.3,6.3]$   & 4.3  & 1.5	& 0.048	& 0.19
& 6.7 \\
$10^{1.5}$ 	& 0.0061 & $[-12.5,12.5]$ & 0.96 & 0.7	& 0.018	&
0.037
& 29.2 \\
\enddata
\tablecomments{Col. (1): Vorticity parameter. Col. (2): Grid spacing in 
Bondi radii. Col. (3): Size of simulation region in the $y$
direction. Col. (4): Run duration in Bondi times. Col. (5):
Approximate time the accretion rate reaches equilibrium, in Bondi
times. Col. (6): Mean accretion rate, in units
of the Bondi rate. Col. (7): Standard deviation in accretion rate, as
a fraction of the mean accretion rate. Col. (8): Mean dimensionless
circulation.
}
\end{deluxetable}

\clearpage

\clearpage

\clearpage

\clearpage

\clearpage

\clearpage

\clearpage

\clearpage

\end{document}